\documentclass[runningheads]{llncs}
\usepackage{url}
\usepackage{color}
\usepackage{hyphenat}
\usepackage{graphicx}
\usepackage{marvosym}
\usepackage{amsmath}
\usepackage[paperheight=235mm,paperwidth=155mm,textwidth=12.2cm,textheight=19.3cm]{geometry}
\usepackage{listings}
\usepackage{float}
\usepackage{url}

\usepackage{amssymb}
\usepackage{comment}
\usepackage{xspace,framed}
\usepackage[ruled,vlined]{algorithm2e}
\usepackage{algorithmic}
\usepackage{soul}
\usepackage{times}
\usepackage{url}
\usepackage[perpage]{footmisc}
\usepackage[dvipsnames]{xcolor}

\usepackage{cite}
\usepackage{multirow}

\makeatletter
\RequirePackage[bookmarks,unicode,colorlinks=true]{hyperref}%
   \def\@citecolor{blue}%
   \def\@urlcolor{blue}%
   \def\@linkcolor{blue}%

\makeatother

\newcommand{\fb}{\textit{FuSeBMC}}

\newcommand{\cou}[1]{\texttt{#1}}

\begin{document}

\title{\fb{} AI: Acceleration of Hybrid Approach through Machine Learning}
\subtitle{(Competition Contribution)}

\author{Kaled M. Alshmrany\textsuperscript{(\Letter)} \inst{1,2}\orcidID{0000-0002-5822-5435}  \and Mohannad Aldughaim \inst{2,3}\orcidID{0000-0002-5822-5435}
     \and
    Chenfeng Wei\inst{2}\orcidID{0009-0008-0416-3006} \and \\ Tom Sweet\inst{4} \and Richard Allmendinger\inst{2} \and Lucas C. Cordeiro\inst{2}\orcidID{0000-0002-6235-4272} }

\institute{
    Institute of Public Administration, Jeddah, Saudi Arabia \and University of Manchester, Manchester, UK \and King Saud University, Riyadh, Saudi Arabia \and SES Escrow, Handforth Cheshire, UK\\ \email{shamranial@ipa.edu.sa}
}

\authorrunning{K. Alshmrany et al.}

\titlerunning{\fb{}-AI: Acceleration of Hybrid Approach through ML}

\maketitle  
\vspace{-0.3cm}
\begin{abstract}
We present \fb{}-AI, a test generation tool grounded in machine learning techniques. \fb{}-AI extracts various features from the program and employs support vector machine and neural network models to predict a hybrid approach's optimal configuration. \fb{}-AI utilizes Bounded Model Checking and Fuzzing as back-end verification engines. \fb{}-AI outperforms the default configuration of the underlying verification engine in certain cases while concurrently diminishing resource consumption.
\end{abstract}

\section{Test-Generation Approach}
\label{sec:overview}
\vspace{-0.2cm}

The success of Machine Learning (ML) 
in automating diverse software engineering tasks is noteworthy, given the escalating complexity of modern software systems~\cite{rossi2023towards}.
A hybrid approach of multiple techniques, including fuzzing, bounded model checking, and abstract interpretation, has proven effective in verifying software compliance with specified requirements~\cite{alshmrany2022fusebmc_FAC}. However, challenges arise, particularly in software with intricate conditions or loops, where the primary obstacle lies in navigating the exponentially expanding program state space and managing resource consumption. Various efforts have been undertaken to enhance the hybrid approach, exemplified by initiatives such as \fb{} Interval Analysis~\cite{fusebmcia} and Tracer~\cite{alshmrany2022fusebmc_FAC}. \fb{}~\cite{alshmrany2021fusebmc,alshmrany2022fusebmc} works as a test generator that synthesizes ``smart seeds'' with properties to enhance the efficiency of its hybrid fuzzer, achieving extensive coverage of programs.
%
To address challenges related to program state explosion and resource usage, \fb{} provides the option of execution with diverse parameters (flags). Unfortunately, determining the optimal flags for a specific program requires expert knowledge, often leading to the execution of hybrid tools with default settings and subsequent compromises in performance.
This paper presents the \fb{}-AI tool to predict the optimal configuration flags for a given program. Specifically, \fb{}-AI employs ML models, support vector machines (SVMs), and neural network (NN) models to predict optimal settings. These ML models undergo training to discern relevant features within the input C program. \fb{}-AI exhibits enhancements in some subcategories in Test-Comp 2024~\cite{beyer2023software}, such as ``ControlFlow'', ``Hardware'', ``Loops'', and ``Software Systems BusyBox MemSafety'', if compared to the default configuration of \fb{}, achieving a $3$\% reduction in resource utilization as reported in Test-Comp 2024. \footnote{\url{https://test-comp.sosy-lab.org/2024/results/results-verified/}}

\section{Software Architecture}
\label{sec:approach}
\begin{figure}
    \includegraphics[width=12cm, height=4.5cm]{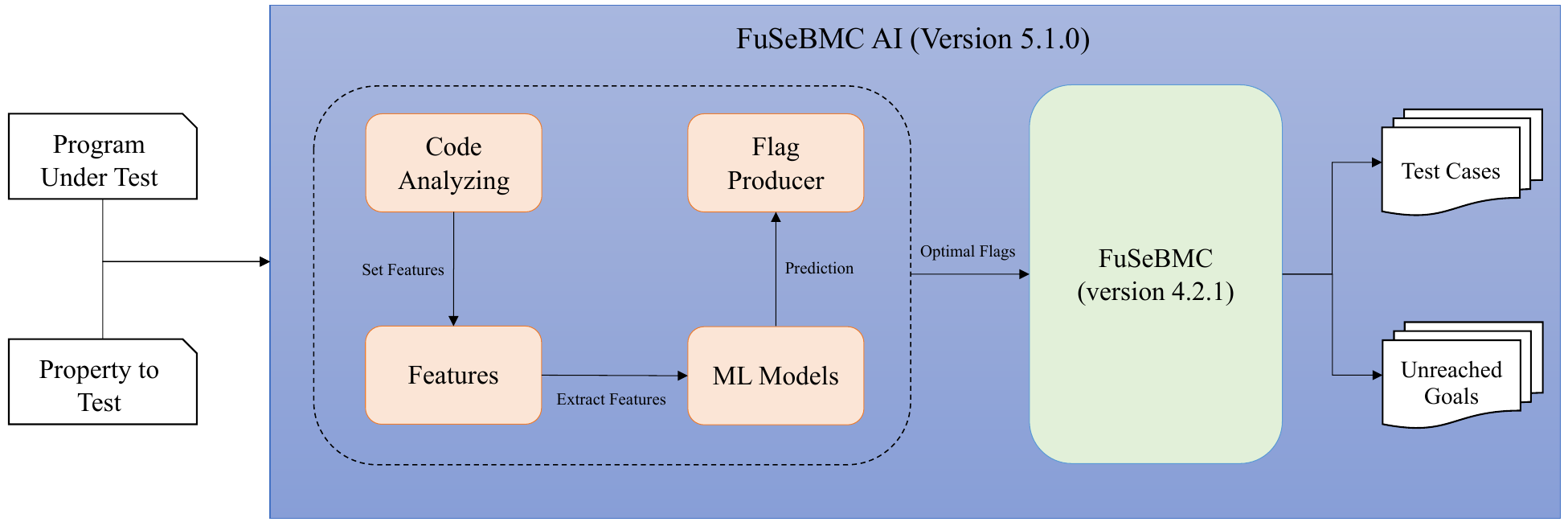}
    \caption{The major components of the \fb{}-AI test generator and how they interact.}
    \label{fig:framework}
\end{figure}

\fb{}-AI builds on top of \fb{}v4.2.1~\cite{alshmrany2021fusebmc,alshmrany2022fusebmc}. The initial step involves analyzing the source code, extracting features that impact training and enhancing the capabilities of \fb{}-AI's engines. Subsequently, these features are stored for future application in ML models. These models, in turn, forecast optimal scores for \fb{}-AI's engines. After that, \fb{}-AI executes the target program using the recommended configuration. Fig.\ref{fig:framework} illustrates the \fb{}-AI framework.

\vspace{-6mm}
\vspace{-3mm}

\subsubsection{Setting Features.}
\label{Setting_Features}
We focus on discerning the features whose values could impact the efficacy and limitations of the engine's performance. This emphasis arose from recognizing that certain programs need specific values for effective handling, particularly those involving arrays and loops. We analyze the Program Under Test (PUT) and extract the features that \fb{}-AI prioritized, which are based on determining the optimal flags and values that could be supplied to the engines of \fb{}-AI (Tab.~\ref{tab:tables}).

\begin{table}[]
\begin{tabular}{|l|l|lll}
\cline{1-2} \cline{4-5}
\textbf{\begin{tabular}[c]{@{}l@{}}Program\\      Features\end{tabular}}              & \textbf{\begin{tabular}[c]{@{}l@{}}Sub\\      Features\end{tabular}}                                                                                                              & \multicolumn{1}{l|}{} & \multicolumn{1}{l|}{\textbf{Flags}}                      & \multicolumn{1}{l|}{\textbf{Values}}                                                                                                                  \\ \cline{1-2} \cline{4-5} 
\multirow{3}{*}{For Loops}                                                            & \multirow{3}{*}{\begin{tabular}[c]{@{}l@{}}For count, For max   depth,\\      For depth avg\end{tabular}}                                                                         & \multicolumn{1}{l|}{} & \multicolumn{1}{l|}{Strategy}                            & \multicolumn{1}{l|}{incr, kinducti}                                                                                                                   \\ \cline{4-5} 
                                                                                      &                                                                                                                                                                                   & \multicolumn{1}{l|}{} & \multicolumn{1}{l|}{Solver}                              & \multicolumn{1}{l|}{boolector, z3}                                                                                                                    \\ \cline{4-5} 
                                                                                      &                                                                                                                                                                                   & \multicolumn{1}{l|}{} & \multicolumn{1}{l|}{Encoding}                            & \multicolumn{1}{l|}{floatbv, fixedbv}                                                                                                                 \\ \cline{1-2} \cline{4-5} 
\multirow{3}{*}{While Loop}                                                           & \multirow{3}{*}{\begin{tabular}[c]{@{}l@{}}While count, While   infinite count\\      While max depth, While depth avg,\\      While infinite with NonDetCall count\end{tabular}} & \multicolumn{1}{l|}{} & \multicolumn{1}{l|}{KStep}                               & \multicolumn{1}{l|}{{[}1,2,3{]}}                                                                                                                      \\ \cline{4-5} 
                                                                                      &                                                                                                                                                                                   & \multicolumn{1}{l|}{} & \multicolumn{1}{l|}{ContextBound}                        & \multicolumn{1}{l|}{{[}2,4{]}}                                                                                                                        \\ \cline{4-5} 
                                                                                      &                                                                                                                                                                                   & \multicolumn{1}{l|}{} & \multicolumn{1}{l|}{Unwind}                              & \multicolumn{1}{l|}{{[}10, -1{]} \#-1 default}                                                                                                        \\ \cline{1-2} \cline{4-5} 
\multirow{2}{*}{Do Loop}                                                              & \multirow{2}{*}{\begin{tabular}[c]{@{}l@{}}Do Count, Do max   depth, \\      Do depth avg, Do infinite count\end{tabular}}                                                        & \multicolumn{1}{l|}{} & \multicolumn{1}{l|}{Fuzz1Enabled}                        & \multicolumn{1}{l|}{{[}0,1{]}}                                                                                                                        \\ \cline{4-5} 
                                                                                      &                                                                                                                                                                                   & \multicolumn{1}{l|}{} & \multicolumn{1}{l|}{\multirow{2}{*}{Fuzz1Time}}          & \multicolumn{1}{l|}{\multirow{2}{*}{\begin{tabular}[c]{@{}l@{}}{[}25,83,188{]} for 250   seconds,\\      (300 - 50) 75\% ,33.3\% ,10\%\end{tabular}}} \\ \cline{1-2}
\multirow{3}{*}{\begin{tabular}[c]{@{}l@{}}If   – Else\\      condition\end{tabular}} & \multirow{3}{*}{\begin{tabular}[c]{@{}l@{}}If count, If max   depth, \\      If depth avg, nested If count,\\      Else count, Else depth avg\end{tabular}}                       & \multicolumn{1}{l|}{} & \multicolumn{1}{l|}{}                                    & \multicolumn{1}{l|}{}                                                                                                                                 \\ \cline{4-5} 
                                                                                      &                                                                                                                                                                                   & \multicolumn{1}{l|}{} & \multicolumn{1}{l|}{\multirow{2}{*}{\textbf{Total run}}} & \multicolumn{1}{l|}{\multirow{2}{*}{\begin{tabular}[c]{@{}l@{}}2*2*2*3*2*2*4 =   384\\      (for each program)\end{tabular}}}                          \\
                                                                                      &                                                                                                                                                                                   & \multicolumn{1}{l|}{} & \multicolumn{1}{l|}{}                                    & \multicolumn{1}{l|}{}                                                                                                                                 \\ \cline{1-2} \cline{4-5} 
\multirow{3}{*}{NonDetCall}                                                           & \multirow{3}{*}{\begin{tabular}[c]{@{}l@{}}Non DetCall   count,\\      Non DetCall depth avg, \\      has Non DetCall in loop\end{tabular}}                                       &                       &                                                          &                                                                                                                                                       \\
                                                                                      &                                                                                                                                                                                   &                       &                                                          &                                                                                                                                                       \\
                                                                                      &                                                                                                                                                                                   &                       &                                                          &                                                                                                                                                       \\ \cline{1-2}
\end{tabular}
    \caption{presents the features that \fb{}-AI prioritized, along with illustrative examples of flags that could be supplied to the engines of \fb{}-AI.}
    \label{tab:tables}
\end{table}
\vspace{-3mm}
\subsubsection{Dataset.}
The SV-Comp benchmarks were selected as the dataset for our training and testing phases for ML models. Our emphasis was on diversity, considering various scenarios, and minimizing repetition to enhance the precision of our approach. We addressed multiple categories: ``no-overflow'', ``termination'', ``unreach-call'', and ``valid-memsafety''\footnote{\url{https://doi.org/10.5281/zenodo.10458701}}. However, for the Test-Comp 2024, our focus narrowed to ``coverage-error-call'' and ``coverage-branches'' encompassing a total of $3352$ benchmarks. In detail, the training set contains 4\% (111 benchmarks) of the coverage-branches benchmarks and 11\% (67 benchmarks) of the coverage-error benchmarks in Test-Comp 2024.
\vspace{-10mm}
\subsubsection{Training and Testing models.}
We focused on four models: Decision Tree Classification (DTC)~\cite{quinlan1986induction}, Support Vector Classification (SVC)~\cite{cortes1995support}, Neural Network Regression (NNR)~\cite{rumelhart1986learning}, and a multi-model (DTC then SVC then NNR). The training phase was executed, followed by using the aforementioned benchmarks. The four models underwent supervised and guided training, ensuring a balanced approach to mitigate repetition during the training phase. The training process involved teaching the models to predict optimal flags for \fb{}-AI’s engines, thereby assisting these engines in determining the most suitable flag values for each category of programs. The classification of outputs was dedicated to facilitating model training (Tab.~\ref{tab:classes}). The classification process involved categorizing ``Cover-Error'' and ``Cover-Branches''. This categorization was based on the extent of coverage or error detection and the corresponding time duration. Comprehensive testing with $384$ different combinations of flags (for each program) was conducted. Consider the cover branches as an illustrative example to provide a more comprehensive understanding of the scale of the conducted experiments. With $111$ benchmarks within the Cover-Branches category, \fb{} is executed approximately $42,624$ times ($111$ multiplied by $384$). Subsequently, we compile a summary encompassing the verification time and verdict for each of the $55,488$ training samples, categorized into ``Cover Error'' ($12,864$ instances) and ``Cover-Branches'' ($42,624$ instances). These samples are assigned ordinal labels ranging from $0$ to $5$, as per the classification outlined in Tab~\ref{tab:classes}. Lower values within the output class are considered more favorable, indicating swift and accurate verdicts.


\vspace{-4mm}
\begin{table}[]
\begin{tabular}{|l|l|l|llll}
\cline{1-3}
\textbf{Testing Result (Cover-Error)}                   & \textbf{Coverage Result (Cover-Branches)} & \textbf{Class} &  &  & \textbf{} & \textbf{} \\ \cline{1-3}
detect bug \& IF restTimeRatio \textgreater{}= 0.8      & score coverage \textgreater{}= 0.85       & 0              &  &  &           &           \\ \cline{1-3}
detect bug \& ELSE IF restTimeRatio \textgreater{}= 0.6 & score coverage \textgreater{}= 0.68       & 1              &  &  &           &           \\ \cline{1-3}
detect bug \& ELSE IF restTimeRatio \textgreater{}= 0.4 & score coverage \textgreater{}= 0.51       & 2              &  &  &           &           \\ \cline{1-3}
detect bug \& ELSE IF restTimeRatio \textgreater{}= 0.2 & score coverage \textgreater{}= 0.34       & 3              &  &  &           &           \\ \cline{1-3}
detect bug \& ELSE IF restTimeRatio \textgreater{}= 0.0 & score coverage \textgreater{}= 0.17       & 4              &  &  &           &           \\ \cline{1-3}
Unknown                                                 & score coverage \textgreater{}= 0.0        & 5              &  &  &           &           \\ \cline{1-3}
\end{tabular}
    \caption{The classification process for ``Cover-Error'' and ``Cover-Branches.''}
    \label{tab:classes}
\end{table}
\vspace{-3mm}
\vspace{-9mm}

\subsubsection{Machine Learning Models.}

DTC, SVC, and NNR models undergo supervised training using the Scikit-learn library~\cite{pedregosa2011scikit}. Each sample is weighted based on class frequency to address class imbalances within the training set. These ML models are trained to predict the output class from $0$ to $5$, considering the features of the Program Under Test (PUT) and a specific set of flags. The trained ML models are then employed to predict the optimal set of flags for \fb{}-AI. Specifically, all $384$ possible flag combinations are tested, and the one resulting in the lowest output class is selected. Due to the computational efficiency of these models, this process is executed rapidly, typically concluding within a few seconds.
\vspace{-3mm}
\section{Strengths and Weaknesses}
\label{sec:strengths-weaknesses}
\vspace{-1mm}

Our proposed hybrid approach demonstrates efficacy in identifying vulnerabilities and attains optimization of computational resources through the AI-based optimisation of fuzzer invocation. An AI model is trained on the classified database, generating optimal flags for executing \fb{}-AI. For instance, the state system generated during the BMC process can be minimized by correctly configuring the unwinding times guided by the training models. This strategic adjustment results in a reduction of traversal spaces and smart seeds generation time. The impact of these enhancements is discernible in the benchmark sets of ``ControlFlow,'' ``Hardware,'' ``Loops,'' ``Software Systems BusyBox MemSafety,'' and ``Termination-Main ControlFlow'' when compared with the default \fb{}. In detail, \fb{}-AI successfully preserved the test case quality in ``ControlFlow'' and ``XCSP'' within the Cover-Error category, achieving a reduction in resource consumption of approximately 86\% in ``ControlFlow'' and 41\% in ``XCSP''. In the Cover-Branches category, \fb{}-AI demonstrated increased coverage by 4\% in ``Software Systems BusyBox MemSafety'' and 1.2\% in ``Hardware''. Furthermore, it achieved an 84\% reduction in resource consumption in ``Termination-Main ControlFlow'' while maintaining the same coverage as the default \fb{}. However, our approach still exhibits limitations. Due to the training process primarily relying on the code sourced from SV-Comp 2023 and Test-Comp 2023, there is an insufficiency in the number and program structures of the training samples. Our ongoing efforts are directed towards addressing this limitation by enriching the training dataset and extending the application of our methodology to open-source software projects.
\vspace{-3mm}
\section{Tool Setup and Configuration}

\fb{}-AI can be used via the python wrapper \texttt{fusebmc.py} to simplify its usage for the competition. Please refer to its help message (\texttt{-h}) for usage instructions. This wrapper runs the \fb{}-AI executable with command line options specific to each supported property. Also, \fb{}-AI offers a graphical user interface (GUI) for enhanced usability \footnote{\url{https://doi.org/10.5281/zenodo.10458701}}.




\vspace{-3mm}
\section{Software Project}

\fb{}-AI is publicly available under the terms of the MIT License at GitHub.\footnote{\url{https://github.com/kaled-alshmrany/FuSeBMC/tree/FuSeBMC-AI}} \fb{}-AI (version 5.1.0) dependencies and instructions for building from source code are all listed in the \cou{README.md} file.

\section{Data-Availability Statement}

All files necessary to run the tool are available on Zenodo~\cite{alshmrany2024fusebmc_ai}.

\vfill

{\small\medskip\noindent{\bf Open Access} This chapter is licensed under the terms of the Creative Commons\break Attribution 4.0 International License (\url{http://creativecommons.org/licenses/by/4.0/}), which permits use, sharing, adaptation, distribution and reproduction in any medium or format, as long as you give appropriate credit to the original author(s) and the source, provide a link to the Creative Commons license and indicate if changes were made.}

{\small \spaceskip .28em plus .1em minus .1em The images or other
third party material in this chapter are included in the\break
chapter's Creative Commons license, unless indicated otherwise in a
credit line to the\break material.~If material is not included in
the chapter's Creative Commons license and\break your intended use
is not permitted by statutory regulation or exceeds the
permitted\break use, you will need to obtain permission directly
from the copyright holder.}


\end{document}